\documentclass[twocolumn,showpacs,pra,amsmath,noshowpacs]{revtex4}
\usepackage{epsfig,graphicx,bm}
\usepackage{color}

\def\be{\begin{equation}}
\def\ee{\end{equation}}
\def\e#1{\label{#1}\end{equation}}
\def\bea{\begin{eqnarray}}
\def\eea{\end{eqnarray}}
\def\ea#1{\label{#1}\end{eqnarray}}
\def\bes#1{\begin{subequations}\label{#1}}
\def\ese{\end{subequations}}

\begin{document}
\title{Notes on squeezed states in $x$-space representation}
\author{Alexander N.\ Korotkov}
\affiliation{Department of Electrical and Computer Engineering, University of
California, Riverside, California 92521}
\date{\today}

\begin{abstract}
In these notes, we discuss squeezed states using the elementary quantum language based on one-dimensional Schr\"odinger equation. No operators are used. The language of quantum optics is mentioned only for a hint  to solve a differential equation. Sections II and III (squeezed vacuum and pure squeezed states) can be used in a standard undergraduate course on quantum mechanics after discussion of a harmonic oscillator. Section IV discusses density matrix of mixed squeezed states in $x$-representation. These notes present explicit (therefore somewhat lengthy) step-by-step derivations. These notes are not intended for publication as a journal paper. 
    \end{abstract}
  \maketitle

\section{Introduction}

In quantum optics \cite{Walls-Milburn-book, Gerry-Knight, Scully-Zubairy, Gardiner-book}, squeezed states are usually introduced via the squeezing operator and discussed in the phase space, with the formalism based on creation and annihilation operators. Here we discuss one-dimensional squeezed states in the $x$-representation, using the formalism based only on elementary quantum mechanics \cite{Griffiths-book, Sakurai-book}.

\section{Squeezed vacuum}

Let us consider a harmonic oscillator with mass $m$ and frequency $\omega$, so the potential energy is
    \be
V(x)=\frac{1}{2}\,m\omega^2x^2.
    \ee
The ground state of this oscillator is \cite{Griffiths-book, Sakurai-book}
    \be
\psi_{\rm gr}(x,t) = \frac{e^{-i\omega t/2}}{\sqrt[4]{2\pi\sigma_{\rm gr}^2}}\, \exp \left( -\frac{x^2}{4\sigma_{\rm gr}^2}\right), \,\,\, \sigma_{\rm gr}^2=\frac{\hbar}{2m\omega},
    \ee
where $\sigma_{\rm gr}^2$ is the $x$-space variance of the ground state probability distribution $|\psi_{\rm gr}(t)|^2$.

Suppose we have prepared an initial state $\psi(x,0)= \exp(-x^2/4D)/{\rm Norm}$. What is the further evolution $\psi(x,t)$ of this state? If $D=\sigma_{\rm gr}^2$, then we have prepared the ground state and therefore it will not evolve (except for the trivial time-factor $e^{-i\omega t/2}$ due to the ground state energy $\hbar\omega/2$). If $D\neq \sigma_{\rm gr}^2$, then this state is called a {\it squeezed vacuum} (especially if $D < \sigma_{\rm gr}^2$). Our goal in this section is to find the explicit time-dependence $\psi(x,t)$ for the squeezed vacuum in $x$-space.

\vspace{0.3cm}

In quantum optics language, the time-dependence of the squeezed vacuum corresponds to the rotation of quadrature axes with frequency $\omega$. From the picture of an ellipse representing the squeezed vacuum in optics, we know that $\psi(x,t)$ should be periodic  with frequency $2\omega$ (except for an accumulating phase). And we also know that a Gaussian state should remain Gaussian with rotation of axes. Also, since $\psi(x,0)$ is symmetric (even), this symmetry should remain in time.

Therefore, the squeezed vacuum should be
    \be
    \psi (x,t) = \frac{1}{\sqrt[4]{2\pi\sigma_{\rm gr}^2 A(t)}} \, \exp\left(-\frac{x^2 [1+iB(t)]}{4\sigma_{\rm gr}^2 A(t)} \right) e^{-i\varphi (t)},
    \label{psi-form}\ee
where $A(t)$, $B(t)$, and $\varphi (t)$ are real dimensionless functions of time, with $A(t)$ and $B(t)$ being $2\omega$-periodic (note that this state is already normalized). In particular, for the ground state $A(t)=1$, $B(t)=0$, and $\varphi(t)=\omega t/2$. Let us find $A(t)$, $B(t)$, and $\varphi (t)$ for the squeezed vacuum.

To use the Schr\"odinger equation \cite{Griffiths-book, Sakurai-book},
    \be
i\hbar \, \frac{\partial \psi}{\partial t}=-\frac{\hbar^2}{2m}\, \frac{\partial^2 \psi}{dx^2}+ \frac{1}{2}\,m\omega^2x^2 \psi,
    \ee
we need to calculate $\partial \psi/\partial t$ and $\partial^2 \psi/dx^2$. This can be easily done for the wavefunction (\ref{psi-form}),
    \begin{eqnarray}
&& \frac{\partial \psi}{\partial t}= \left[-\frac{\dot{A}}{4A} +\frac{x^2 (1+iB)\dot{A}}{4\sigma_{\rm gr}^2 A^2} -\frac{ix^2 \dot{B}}{4\sigma_{\rm gr}^2 A} -i\dot{\varphi} \right] \psi , \qquad
    \\
&& \frac{\partial ^2\psi}{\partial x^2}=\left[ \left(-\frac{x(1+iB)}{2\sigma_{\rm gr}^2 A}\right)^2 - \frac{1+iB}{2\sigma_{\rm gr}^2 A} \right] \psi .
    \end{eqnarray}
Substituting these derivatives into the Schr\"odinger equation, we obtain
    \begin{eqnarray}
&& i\hbar \left[-\frac{\dot{A}}{4A} +\frac{x^2 (1+iB)\dot{A}}{4\sigma_{\rm gr}^2 A^2} -\frac{ix^2 \dot{B}}{4\sigma_{\rm gr}^2 A} -i\dot{\varphi} \right] =
\nonumber \\
&& \hspace{0.5cm} = -\frac{\hbar^2}{2m} \left[ \frac{x^2(1+iB)^2}{4\sigma_{\rm gr}^4 A^2} - \frac{1+iB}{2\sigma_{\rm gr}^2 A} \right] +\frac{1}{2} m\omega^2 x^2 , \qquad
    \end{eqnarray}
which is the same as
    \begin{eqnarray}
&& -\frac{\dot{A}}{A} +\frac{x^2 (1+iB)\dot{A}}{\sigma_{\rm gr}^2 A^2} -\frac{ix^2 \dot{B}}{\sigma_{\rm gr}^2 A} -4i\dot{\varphi}  =
\nonumber \\
&& \hspace{0.5cm} = \frac{i\hbar}{2m} \left[ \frac{x^2(1+iB)^2}{\sigma_{\rm gr}^4 A^2} - 2\frac{1+iB}{\sigma_{\rm gr}^2 A} \right] -\frac{2i}{\hbar} m\omega^2 x^2 . \qquad
    \label{psi-dif-eq}\end{eqnarray}

Since this relation is valid for any $x$, it gives us two equations: for $x^2$-tems and for $x^0$-terms. Let us start with the equation for $x^2$-component (multiplied by $\sigma_{\rm gr}^2$),
    \be
\frac{(1+iB)\dot{A}}{A^2} -\frac{i \dot{B}}{A}  =
  \frac{i\hbar (1+iB)^2}{2m\sigma_{\rm gr}^2 A^2} -\frac{2i}{\hbar} m\omega^2 \sigma_{\rm gr}^2.
    \label{psi-x2-component}\ee
Using  $\sigma_{\rm gr}^2=\hbar/(2m\omega)$ and multiplying all terms by $A^2$, we can rewrite Eq.\ (\ref{psi-x2-component}) as
    \be
    (1+iB)\dot{A} -i A \dot{B} = i\omega(1+iB)^2 -i\omega A^2.
    \ee
Since $A$ and $B$ are real, this equation gives us two equations: for the real and imaginary components,
    \begin{eqnarray}
&& \dot{A}=-2\omega B,
    \label{dot-A} \\
&& B\dot{A} -A\dot{B}= \omega (1-B^2-A^2).
    \label{B-A-dot}\end{eqnarray}

These two equations are sufficient to find $A(t)$ and $B(t)$. We will discuss the solution a little later; before that, let us consider the $x^0$-component of Eq.\  (\ref{psi-dif-eq}), which is
    \be
-\frac{\dot{A}}{A} -4i\dot{\varphi}  = -\frac{i\hbar}{m} \frac{1+iB}{\sigma_{\rm gr}^2 A} = \frac{2\omega}{A}\, (-i+B).
    \ee
The real part of this equation gives $\dot{A}=-2\omega B$, which is the same as Eq.\ (\ref{dot-A}). The imaginary part gives
    \be
    \dot{\varphi}=\frac{\omega}{2A},
    \label{dot-varphi}\ee
from which we can find $\varphi (t)$ if we know $A(t)$.

Thus, we only have to solve Eqs.\ (\ref{dot-A}) and (\ref{B-A-dot}). Since we know that both $A(t)$ and $B(t)$ should be periodic with frequency $2\omega$, it is quite simple to guess the solution:
    \begin{eqnarray}
&&    A(t)= A_0 +\Delta A \cos (2\omega t +\phi_{\rm sq}),
    \label{A(t)}\\
&& B(t)=\Delta A \sin (2\omega t +\phi_{\rm sq}),
    \label{B(t)}\end{eqnarray}
where $\phi_{\rm sq}$ is the initial phase for oscillations of squeezing and $A_0>\Delta A\geq 0$. It is easy to see that Eq.\ (\ref{dot-A}) is satisfied for any $A_0$, $\Delta A$, and $\phi_{\rm sq}$, while to satisfy Eq.\ (\ref{B-A-dot}) we need the condition
    \be
    (A_0+\Delta A)(A_0-\Delta A)=1,
    \label{A-0-product}\ee
which is a familiar condition for the product of minimum and maximum quadrature variances of a squeezed state. Note that for the initial state discussed at the beginning, we need $\phi_{\rm sq}=n\pi$ with an integer $n$. Also note that the squeezed vacuum (\ref{psi-form}) is a minimum-uncertainty state, $\sigma_x\sigma_p=\hbar/2$, only when $2\omega t+\phi_{\rm sq}=n \pi$ with integer $n$, i.e., four times per oscillator period.

Thus, we have found the evolution of the squeezed vacuum (\ref{psi-form}) up to the overall phase $\varphi (t)$, which is not quite important and can be found via Eq.\ (\ref{dot-varphi}). It is interesting that there is an explicit solution,
    \begin{eqnarray}
&& \hspace{-0.3cm} \varphi (t) = \frac{\omega}{2} \int^t \frac{dt'}{A_0 +\Delta A \cos (2\omega t' +\phi_{\rm sq})}
    \\
&& \hspace{0.5cm} = \frac{1}{2}\, {\rm atan}\left[ (A_0-\Delta A) \, {\rm tan}(\omega t+\frac{\phi_{\rm sq}}{2})\right] +{\rm const}, \qquad
    \label{varphi-explicit}\end{eqnarray}
where each discontinuity of the tangent should be compensated by an increase of the constant by $\pi/2$. Note that in deriving Eq.\ (\ref{varphi-explicit}) we have used Eq.\ (\ref{A-0-product}). It is easy to see that for each $\Delta t=\pi/\omega$ (half-period), the phase increases by $\Delta \varphi = \pi/2$, so that $\Delta \varphi/\Delta t=\omega/2$, which is the same value as for the ground state.

\section{Pure squeezed states}

Now let us generalize the $x$-space evolution of the squeezed vacuum to the $x$-space evolution of an arbitrary (pure) squeezed state. Since we know that for a squeezed state the evolution of the center is decoupled from squeezing, it is natural to guess the following wavefunction for a squeezed state:
    \begin{eqnarray}
&&    \psi (x,t) = \frac{1}{\sqrt[4]{2\pi\sigma_{\rm gr}^2 A}} \, \exp\left[ -\frac{(x-x_{\rm c})^2 (1+iB)}{4\sigma_{\rm gr}^2 A} \right]
    \nonumber \\
&& \hspace{1.2cm} \times \,  \exp (ip_{\rm c} x/\hbar) \, e^{-i\varphi },
    \label{psi-gen-form}\end{eqnarray}
where $A(t)$ and $B(t)$ are the same functions of time as for the squeezed vacuum [given by Eqs.\ (\ref{A(t)})--(\ref{A-0-product})], while the overall phase $\varphi(t)$ is now different, and the state center evolves classically,
    \begin{eqnarray}
&& x_{\rm c}= X_{\rm amp} \cos (\omega t +\phi_{\rm c}) ,
    \label{x-c}\\
&& p_{\rm c}= m\dot{x}_{\rm c}=-m\omega X_{\rm amp}  \sin (\omega t +\phi_{\rm c}),
    \label{p-c}\end{eqnarray}
where $X_{\rm amp}$ is the amplitude of classical oscillations of the center and the initial phase $\phi_{\rm c}$ is not related to the initial phase $\phi_{\rm sq}$ of squeezing oscillations. Let us show that the time-dependence of an arbitrary squeezed state given by Eqs.\ (\ref{psi-gen-form})--(\ref{p-c}) and (\ref{A(t)})--(\ref{A-0-product}) satisfy the Schr\"odinger equation.

The derivatives of $\psi(x,t)$ given by Eq.\ (\ref{psi-gen-form}) are
    \begin{eqnarray}
&& \hspace{-0.3cm} \frac{\partial \psi}{\partial t}= \Bigg[ -\frac{\dot{A}}{4A} +\frac{(x-x_{\rm c})^2 (1+iB)\dot{A}}{4\sigma_{\rm gr}^2 A^2} -\frac{i(x-x_{\rm c})^2 \dot{B}}{4\sigma_{\rm gr}^2 A}
    \nonumber \\
&& \hspace{1.0cm} + \, \frac{(x-x_{\rm c})\dot{x}_{\rm c} (1+iB)}{2\sigma_{\rm gr}^2 A} +\frac{i\dot{p}_{\rm c} x}{\hbar} -i\dot{\varphi} \Bigg] \psi , \qquad
    \\
&& \hspace{-0.3cm}  \frac{\partial ^2\psi}{\partial x^2}=\left[ \left[-\frac{(x-x_{\rm c})(1+iB)}{2\sigma_{\rm gr}^2 A} +\frac{ip_{\rm c}}{\hbar}\right]^2 - \frac{1+iB}{2\sigma_{\rm gr}^2 A} \right] \psi  , \qquad
    \end{eqnarray}
so the Schr\"odinger equation is [instead of Eq.\ (\ref{psi-dif-eq})]
    \begin{eqnarray}
&& -\frac{\dot{A}}{A} +\frac{(x-x_{\rm c})^2 (1+iB)\dot{A}}{\sigma_{\rm gr}^2 A^2} -\frac{i(x-x_{\rm c})^2 \dot{B}}{\sigma_{\rm gr}^2 A} -4i\dot{\varphi}
    \nonumber \\
&&\hspace{1cm} +\frac{2(x-x_{\rm c}) \dot{x}_{\rm c}(1+iB)}{\sigma_{\rm gr}^2 A}+4i \, \frac{\dot{p}_{\rm c} x}{\hbar} =
\nonumber \\
&& \hspace{0.0cm} = \frac{i\hbar}{2m} \bigg[  \frac{(x-x_{\rm c})^2(1+iB)^2}{\sigma_{\rm gr}^4 A^2} - 2\frac{1+iB}{\sigma_{\rm gr}^2 A} -4\frac{p_{\rm c}^2}{\hbar^2}
    \nonumber \\
&&\hspace{1cm} -4i\frac{p_{\rm c}}{\hbar} \frac{(x-x_{\rm c})(1+iB)}{\sigma_{\rm gr}^2 A} \bigg] -\frac{2i}{\hbar} m\omega^2 x^2 . \qquad
    \label{psi-gen-dif-eq}\end{eqnarray}

This equation now gives three equations: for $x^2$, $x^1$, and $x^0$ components. It is easy to check that the $x^2$-component still gives Eq.\ (\ref{psi-x2-component}). Therefore, Eqs.\ (\ref{dot-A})--(\ref{B-A-dot}) for $A(t)$ and $B(t)$ and correspondingly their solutions, Eqs.\ (\ref{A(t)})--(\ref{A-0-product}), are still valid for an arbitrary squeezed state. The $x^1$-component of Eq.\ (\ref{psi-gen-dif-eq}) multiplied by $\sigma_{\rm gr}^2$ gives
    \begin{eqnarray}
&& \frac{-2x_{\rm c} (1+iB)\dot{A}}{A^2} +\frac{2ix_{\rm c} \dot{B}}{A} +\frac{2 \dot{x}_{\rm c}(1+iB)}{A}+4i \, \frac{\dot{p}_{\rm c}}{\hbar}\, \sigma_{\rm gr}^2  =
\nonumber \\
&& \hspace{1.0cm} = \frac{i\hbar}{2m} \,  \frac{-2x_{\rm c}(1+iB)^2}{\sigma_{\rm gr}^2 A^2} + \frac{2p_{\rm c}}{m} \frac{1+iB}{A} . \qquad
    \label{x-1}\end{eqnarray}
Note that several terms here are the same as the terms in Eq.\ (\ref{psi-x2-component}) multiplied by $-2x_{\rm c}$. Subtracting Eq.\ (\ref{psi-x2-component}) multiplied by $-2x_{\rm c}$ from Eq.\ (\ref{x-1}) and using $\sigma_{\rm gr}^2=\hbar/2m\omega$, we obtain
    \be
\frac{2 \dot{x}_{\rm c}(1+iB)}{A}+2i \, \frac{\dot{p}_{\rm c}}{m\omega}  =
\frac{2p_{\rm c}}{m} \frac{1+iB}{A} - 2ix_{\rm c} \omega  ,
    \label{x-1-diff}\ee
which is obviously satisfied for $x_{\rm c}$ and $p_{\rm c}$ given by Eqs.\ (\ref{x-c}) and (\ref{p-c}). Thus, the $x^1$-component of Eq.\ (\ref{psi-gen-dif-eq}) is satisfied. Finally, the $x^0$-component gives
    \begin{eqnarray}
&&  \hspace{-0.3cm} -\frac{\dot{A}}{A} +\frac{x_{\rm c}^2 (1+iB)\dot{A}}{\sigma_{\rm gr}^2 A^2} -\frac{ix_{\rm c}^2 \dot{B}}{\sigma_{\rm gr}^2 A} -4i\dot{\varphi}
  -\frac{2x_{\rm c} \dot{x}_{\rm c}(1+iB)}{\sigma_{\rm gr}^2 A} =
\nonumber \\
&& \hspace{-0.3cm} \frac{i\hbar}{2m} \bigg[  \frac{x_{\rm c}^2(1+iB)^2}{\sigma_{\rm gr}^4 A^2} - 2\frac{1+iB}{\sigma_{\rm gr}^2 A} -4\frac{p_{\rm c}^2}{\hbar^2}
    +4i\frac{p_{\rm c}}{\hbar} \frac{x_{\rm c}(1+iB)}{\sigma_{\rm gr}^2 A} \bigg] . \qquad
    \label{x-0-gen}\end{eqnarray}
The real part of this equation is
    \be
  \hspace{-0.1cm} -\frac{\dot{A}}{A} +\frac{x_{\rm c}^2 \dot{A}}{\sigma_{\rm gr}^2 A^2}
  -\frac{2x_{\rm c} \dot{x}_{\rm c}}{\sigma_{\rm gr}^2 A} =  \frac{\hbar B}{m\sigma_{\rm gr}^2 A}
  -\frac{\hbar x_{\rm c}^2B}{m \sigma_{\rm gr}^4 A^2}     -2\frac{p_{\rm c}x_{\rm c}}{m\sigma_{\rm gr}^2 A} .
    \label{x-0-gen-Re}\ee
The first terms on both sides are equal to each other because $\dot{A}=-2\omega B$ [see Eq.\ (\ref{dot-A})]. The second terms are equal  because of the same reason. The third terms are equal because $\dot{x}_{\rm c}=p_{\rm c}/m$. So, Eq.\ (\ref{x-0-gen-Re}) is satisfied. Finally, the imaginary part of Eq.\ (\ref{x-0-gen}) gives (after some algebra)
        \begin{eqnarray}
&& \dot{\varphi} = \frac{\omega}{2 A} + \frac{x_{\rm c}^2 B\dot{A}}{4\sigma_{\rm gr}^2 A^2} -\frac{x_{\rm c}^2 \dot{B}}{4\sigma_{\rm gr}^2 A}
 - \frac{\omega x_{\rm c}^2(1-B^2)}{4\sigma_{\rm gr}^2 A^2}
 + \frac{p_{\rm c}^2}{2\hbar m} \quad
    \nonumber \\
&& \hspace{0.4cm} = \frac{\omega}{2 A} + \frac{m\omega^2 (X_{\rm amp}^2 -x_{\rm c}^2)}{2\hbar}  .
    \label{x-0-gen-Im}\end{eqnarray}
It is simple to solve this equation by adding the explicit solution (\ref{varphi-explicit}) for $x_{\rm c}=X_{\rm amp}=0$ and the integral of the last term in Eq.\ (\ref{x-0-gen-Im}).

Thus, we have shown that the general (pure) squeezed state in $x$-space has the form given by Eqs.\ (\ref{psi-gen-form})--(\ref{p-c}), (\ref{A(t)})--(\ref{A-0-product}), and (\ref{x-0-gen-Im}).

\section{Mixed squeezed (Gaussian) states }

Using the wavefunction (\ref{psi-gen-form}), let us construct the density matrix
    \begin{eqnarray}
&& \hspace{-0.4cm} \rho(x,x')=\psi(x)\, \psi^*(x')=\frac{1}{\sqrt{2\pi \sigma_{\rm gr}^2 A}} \, \exp \left[ -\frac{(\frac{x+x'}{2}-x_{\rm c})^2}{2\sigma_{\rm gr}^2 A} \right]
    \nonumber \\
&& \hspace{0.6cm} \times  \, \exp \left[ -\frac{(\frac{x-x'}{2})^2}{2\sigma_{\rm gr}^2 A} \right] \,
\exp\left[ -iB  \frac{(\frac{x+x'}{2}-x_{\rm c})(x-x')}{2\sigma_{\rm gr}^2 A}   \right]
    \nonumber \\
&& \hspace{0.6cm} \times \, \exp\left[\, i\, \frac{p_{\rm c}(x-x')}{\hbar}\right] .
    \label{rho-pure}\end{eqnarray}

Now let us probabilistically mix the states with $x_{\rm c}$ and $p_{\rm c}$ having the Gaussian probability distribution
    \begin{eqnarray}
&& p(x_{\rm c}, p_{\rm c}) = \frac{1}{\sqrt{2\pi \sigma_{\rm a}^2}} \, \exp \left[ -\frac{(x_{\rm c}-\bar{x}_{\rm c})^2}{2\sigma_{\rm a}^2}\right]
    \nonumber \\
&& \hspace{1.4cm}   \times \, \frac{1}{\sqrt{2\pi (m\omega \sigma_{\rm a})^2}} \,
    \exp\left[ -\frac{(p_{\rm c}-\bar{p}_{\rm c})^2}{2(m\omega\sigma_{\rm a})^2} \right] , \qquad
    \label{x-c-distrib}\end{eqnarray}
so that the additional $x_{\rm c}$-spread has variance $\sigma_{\rm a}^2$ and the spread of $p_{\rm c}/m\omega$ has the same variance $\sigma_{\rm a}^2$, while the averaged centers $\bar{x}_{\rm c}$ and $\bar{p}_{\rm c}$ evolve according to Eqs.\ (\ref{x-c}) and (\ref{p-c}), i.e.,
    \be
    \dot{\bar x}_{\rm c}=\bar{p}_{\rm c}/m, \,\,\, \dot{\bar p}_{\rm c} = -m\omega^2 \bar{x}_{\rm c}.
    \label{dot-x-bar}\ee

It is easy to see that the evolution (\ref{dot-x-bar}) of $\bar{x}_{\rm c}$ and $\bar{p}_{\rm c}$ does not change the probability distribution (\ref{x-c-distrib}) for $x_{\rm c}$ and $p_{\rm c}$, which evolve according to Eqs.\ (\ref{x-c})--(\ref{p-c}). The easiest way to see this fact is to consider an evolution on the plane of $x_{\rm c}$ and $p_{\rm c}/m\omega$, so that Eqs.\ (\ref{x-c})--(\ref{p-c}) as well as (\ref{dot-x-bar}) describe a rotation with frequency $\omega$ along a circle, while the Gaussian distribution (\ref{x-c-distrib}) is isotropic. Therefore, if we average the density matrix (\ref{rho-pure}) over $x_{\rm c}$ and $p_{\rm c}$ with the distribution (\ref{x-c-distrib}), the result {\it automatically} satisfies the Schr\"odinger equation.

For averaging the density matrix (\ref{rho-pure})  over $x_{\rm c}$, the following formula is useful:
    \begin{eqnarray}
&&   \int_{-\infty}^\infty \frac{\exp \left[ -\frac{(x+y)^2+a(x+y)}{2\sigma_1^2} \right] }{\sqrt{2\pi\sigma_1^2}} \,\, \frac{ \exp \left(-\frac{y^2}{2\sigma_2^2}\right) }{\sqrt{2\pi\sigma_2^2}} \,\, dy=
    \nonumber \\
&& \hspace{0.5cm} =   \frac{ \exp \left[ -\frac{x^2+ax}{2(\sigma_1^2+\sigma_2^2)} \right] }{\sqrt{2\pi (\sigma_1^2+\sigma_2^2)}}
 \,\,  \exp \left[\frac{a^2 \sigma_2^2}{8 \sigma_1^2 (\sigma_1^2 +\sigma_2^2)  }\right] , \quad
    \end{eqnarray}
where $a$ can be complex. Similarly, for averaging over $p_{\rm c}$ we can use
     \be
  \int_{-\infty}^\infty e^{ a (x+y) } \,
    \frac{\exp (-y^2/2\sigma^2)}{\sqrt{2\pi\sigma^2}} \,  dy= e^{ax } \exp(a^2\sigma^2/2) .
    \ee
Therefore, averaging of the density matrix (\ref{rho-pure}) over the distribution (\ref{x-c-distrib}) gives
    \begin{eqnarray}
&& \hspace{-0.3cm} \rho(x,x')= \frac{1}{\sqrt{2\pi \sigma_\Sigma^2 }} \, \exp \left[ -\frac{(\frac{x+x'}{2}-\bar{x}_{\rm c})^2}{2\sigma_\Sigma^2} \right]
    \nonumber \\
&& \hspace{0.3cm} \times  \,
\exp\left[ -iB  \frac{(\frac{x+x'}{2}-\bar{x}_{\rm c})(x-x')}{2\sigma_\Sigma^2}   \right] \, \exp \left[\, i\, \frac{\bar{p}_{\rm c}(x-x')}{\hbar}\right]
    \nonumber \\
&& \hspace{-0.2cm} \times \,
\exp \left[ -\frac{(\frac{x-x'}{2})^2}{2\sigma_{\rm gr}^2 A}
- \frac{B^2 \sigma_{\rm a}^2(x-x')^2}{8\sigma_{\rm gr}^2 A \sigma_\Sigma^2}-
\frac{\sigma_{\rm a}^2 (x-x')^2}{8\sigma_{\rm gr}^4} \right]  , \,\,\,\, \quad
    \label{rho-mixed}\end{eqnarray}
where $\sigma_\Sigma^2=\sigma_{\rm gr}^2 A+\sigma_{\rm a}^2$. Using Eqs.\ (\ref{A(t)})--(\ref{A-0-product}), the last exponential factor can be expressed as
\be
\exp \left[ -\frac{(\frac{x-x'}{2})^2}{2\sigma_\Sigma^2}  \,\, \frac{\sigma_{\rm gr}^4+\sigma_{\rm a}^4+ 2A_0 \sigma_{\rm gr}^2 \sigma_{\rm a}^2}{\sigma_{\rm gr}^4 } \right] .
\label{exp-term}\ee
We see that if we introduce $\tilde{A}\equiv A+\sigma_{\rm a}^2/\sigma_{\rm gr}^2$ and $\tilde{A}_0 \equiv A_0+\sigma_{\rm a}^2/\sigma_{\rm gr}^2$, then $\sigma_{\Sigma}^2 = \sigma_{\rm gr}^2\tilde{A}$ and the second fraction in  Eq.\ (\ref{exp-term}) can be rewritten as $1+ (\sigma_{\rm a}/\sigma_{\rm gr})^4 + 2A_0 (\sigma_{\rm a}/\sigma_{\rm gr})^2 =P=(\tilde{A}_0 +\Delta A)(\tilde{A}_0 -\Delta A)$.

Therefore, removing tilde signs and  overbars ($\tilde A \to A$, $\tilde A_0 \to A_0$, $\bar{x}_{\rm c}\to x_{\rm c}$, $\bar{p}_{\rm c}\to p_{\rm c}$), we can rewrite the averaged density matrix (\ref{rho-mixed}) as
     \begin{eqnarray}
&& \hspace{-0.4cm} \rho(x,x')=\frac{1}{\sqrt{2\pi \sigma_{\rm gr}^2 A}} \, \exp \left[ -\frac{(\frac{x+x'}{2}-x_{\rm c})^2}{2\sigma_{\rm gr}^2 A} \right]
    \nonumber \\
&& \hspace{0.2cm} \times  \, \exp \left[ -P\,\frac{(\frac{x-x'}{2})^2}{2\sigma_{\rm gr}^2 A} \right] \,
\exp\left[ -iB  \frac{(\frac{x+x'}{2}-x_{\rm c})(x-x')}{2\sigma_{\rm gr}^2 A}   \right]
    \nonumber \\
&& \hspace{0.2cm} \times \, \exp\left[\, i\, \frac{p_{\rm c}(x-x')}{\hbar}\right] ,
    \label{rho-mixed-2}\end{eqnarray}
which differs from Eq.\ (\ref{rho-pure}) for the pure state by {\it only} the factor $P$ in the second exponent. This is the general form of a
Gaussian state of an oscillator. In Eq.\ (\ref{rho-mixed-2}), the time-dependence of the center $(x_{\rm c}, p_{\rm c})$ is classical and is given by Eqs.\ (\ref{x-c}) and (\ref{p-c}); the center oscillates with frequency $\omega$. The parameters $A(t)$ and $B(t)$ are given by Eqs.\ (\ref{A(t)}) and (\ref{B(t)}), they oscillate with frequency $2\omega$. The parameter $P$ is the product of maximum and minimum dimensionless variances,
    \be
    P=A_{\rm max} A_{\rm min}=(A_0+\Delta A)(A_0-\Delta A) \geq 1,
    \label{P-def}\ee
so that $P=1$ corresponds to a pure state, while $P>1$ corresponds to a mixed Gaussian state.

\section{Summary}

In the $x$-space representation, the squeezed vacuum wavefunction is given by Eq.\ (\ref{psi-form}) with dimensionless variance $A(t)$ and phase parameter $B(t)$ given by Eqs.\ (\ref{A(t)})--(\ref{A-0-product}), and the overall phase $\varphi (t)$ given by Eq.\ (\ref{varphi-explicit}). For an oscillator with frequency $\omega$, the shape of the wavefunction oscillates with frequency $2\omega$.

The wavefunction of a pure squeezed state is given by Eq.\ (\ref{psi-gen-form}) with the same parameters $A(t)$ and $B(t)$ as for the squeezed vacuum, while the center position $x_{\rm c}(t)$ and momentum $p_{\rm c}(t)$ are given by Eqs.\ (\ref{x-c}) and (\ref{p-c}). The center oscillates with frequency $\omega$, while the variance oscillates with frequency $2\omega$.

The density matrix of a mixed (Gaussian) squeezed state is given by Eq.\ (\ref{rho-mixed-2}) with the same equations for $A(t)$, $B(t)$, $x_{\rm c}(t)$,  and $p_{\rm c}(t)$, but with an additional parameter $P$ given by  Eq.\ (\ref{P-def}), which replaces Eq.\ (\ref{A-0-product}).

\end{document}